# A PRELIMINAR EVIDENCE OF QUANTUM LIKE BEHAVIOR IN MEASUREMENTS OF MENTAL STATES

Elio Conte [+], Orlando Todarello [-], Antonio Federici [+],

Francesco Vitiello [+], Michele Lopane [-]

[+] Department of Pharmacology and Human Physiology;
TIRES-Center for Innovative Technologies for Signal Detection and Processing,
University of Bari, Italy;

[-] Department of Neurological and Psychiatric Sciences, University of Bari, Italy

and

Andrei Khrennikov [°]

[°] International Center for Mathematical Modeling in Physics and Cognitive Sciences,
MSI, University of Växjö, S-35195, Sweden

**Abstract**

It has been reached experimental evidence that mental states behave in a quantum like, context dependent, manner with indication that mental phenomena cannot be completely embedded into the traditional physical space. Calculations on the quantum like model of mental states are reported in detail.

# 1. INTRODUCTION

In studies on cognitive function there is first of all the problem to define where and how it may be considered the location or, alternatively, the non locality of consciousness as well as of more primitive cognitive processes. The problem may be regarded as a basic question of physics as well as it is widely discussed also in philosophical, neurophysiological and psychological studies.

There is a large variety of views starting with the primary question if and where consciousness is located in human brain. Starting with I. Kant [1], basic contributions derived from L. Bianchi [2], I.P. Pavlov [3], W. Bechterew [4], H. Eichenbaum [5], J.M.D. Fuster [6] and H.and A.R. Damasio [7].

One possible criticism to such philosophic, neurophysiological and psychological positions, all devoted to assign a space of location to consciousness and cognitive functions, is that all such elaboration is based on one assumed principle. Authors admit in principle that the geometry is fixed, namely the Euclidean one, with connected idea of an absolute, primitive and unique space representative of such geometry. As we know, I. Kant started with such assumed principle and the following developments of modern philosophic, neuro physiological and psychological approaches maintained always the conception that the space is the absolute Euclidean space, that this space is primary and that nothing may happen without strong and direct relation to such space. Since space is identified with the absolute, primitive and unique Euclidean space, it remains the problem to identify the place of consciousness and of primitive cognitive processes in this space. As we know, many attempts were performed in order to identify the place of consciousness in such kind of space and, to this regard, many studies were performed in order to reduce consciousness to the dynamics of excited neurons (see e.g. Dynamical Systems Approach [8]-[10]), but, in spite of enormous efforts to find finally the place of consciousness by this way, evidences emerged that it cannot be conceived as located in such traditional physical space.

The problem could be the choice of principle that we made for the geometry. The vision of an existing primary, unique and absolute space may be not adequate in the tentative to find location of consciousness. Natural phenomena, at different scales of their dynamics, could require different geometries allocated in different spaces. One could retain to observe here that the use of various geometries is really made in physics and, in general, in science as it is the case, as example, in special and general relativity.

But we do not speak just about various geometries on physical space. As it was underlined in [11], mental phenomena cannot be completely embedded into physical space. We should introduce additional mental coordinates to describe flows of minds. In series of works [11] there was developed a concrete model of mental space, namely p-adic mental space. In this model mathematical features of p-adic mental coordinates differ strongly from mathematical features of real continuous physical coordinates. The p-adic mental coordinates have treelike structures, they are discontinuous and totally disconnected. Roughly speaking by [11] the mind cannot be embedded into physical space represented by the Cartesian product of straight lines, but it can be embedded into a mental space represented by the Cartesian product of p-adic trees. In [12], [13] the p-adic mental space approach [11] was developed to describe general information spaces. We remark that Chalmers considered (in the philosophical framework) information spaces in [14] as the basis for solving mind-body problem.

However, it might be that both physical and mental coordinates can be represented as images of coordinates of a more fundamental space, the PRESPACE. As it was repeatedly outlined by B. Hiley [15], the problem here is that even in physics the usual space-time must not be considered as primary. He wrote that if this space-time is taken as primary, then, ipso facto, locality becomes absolute and the use of such manifolds becomes unable to solve the problem of non locality for physical systems, of non locality of psychological functions and, in particular, of cognitive states. Indeed, the usual space-time manifold dominates events of classical-deterministic-physics and this science has locality as well as determinism built into it right at the beginning. Our basic insistence to take such given space-time as basic of all the natural dynamics is at fault. The usual space-time could be merely an appearance, a feature that may be abstracted from some deeper structure and that holds as long as our basic paradigmatic approach is based on an accepted deterministic vision of the things while instead in the picture of a more deeper structure it no more can be taken as basic and primary. In this manner locality itself becomes a mere relationship able to appear with convenience in our macroscopic and in classical world but it actually would not be universally valid. Quantum phenomena, actually happening in such deeper structure, would be projected into our usual space-time by our macroscopic instruments of observation and, in general, by our empirical experience.

Thus, the idea of a PRESPACE arises. Some founder fathers of physics as J.A Wheeler [16] and A.S. Eddington [17] outlined the importance of this viewpoint. In principle Chalmers [14] information space can also be considered as a kind of PRESPACE. Recently in [18] PRESPACE was introduced (by pure probabilistic reasons) as a Kolmogorov probability space describing pre quantum world. Such a PRESPACE can be mapped (with huge reduction of information) into quantum-like Hilbert space.

B. Hiley [15] discussed the importance to consider PRESPACE as proper indication of the mental space for location of consciousness and of primitive cognitive processes. In [19] one of us gave proper formal results to this regard. It was suggested that the actual space of Clifford algebra is the deeper structure from which usual space-time derives. The space of the Clifford algebra should be the space and the usual space-time should be its mere projection. In addition, it was shown [19] that the basic Clifford axioms seem to hold for a bare bone skeleton of quantum theory still with possibility to derive, without any other physical assumptions, classical diffusion equation as well as Schrödinger equation from the same basic axiomatic Clifford set. In conclusion, in order to attempt to define a PRESPACE as location for consciousness and primary cognitive processes, we should be able to define that consciousness and cognitive functions move out from the scheme of classical deterministic physics following in consequence a quantum-like behavior. However, also this is an argument that requires to be characterized in detail.

In [20] another author developed so called quantum-like statistical model of cognitive measurements. This model can be considered as a cognitive application of the general quantum-like model [21]. Quantum-like statistical models are characterized by violations of the formula of addition of probabilities of alternatives. There can arise various additional interference terms and, in particular, trigonometric interference. Since the latter interference is one of distinguishing features of the ordinary quantum model describing statistical processes in the micro world, the standard (Bohr-Heisenberg-Schrödinger-Dirac-vonNeumann) quantum model is, of course, quantum-like. But in principle there can exist models in which quantum-like statistical behavior need not be related to "ordinary quantum processes." In [21] there was demonstrated that quantum-like probabilistic behavior is a consequence of contextual high sensitivity of systems (physical, mental, social,…) to changes of contexts (complexes of conditions – e.g. physical, or mental, or social,…).

In [19] there was presented the conjecture that mental systems can exhibit quantum-like statistical behavior. As it was already underlined, such a behavior need not be connected with " ordinary quantum processes" in micro world. Moreover, in the view of our discussion about mental spaces and fundamental PRESPACE quantum-like cognitive behavior need not be induced by processes in physical space-time.

## 2. ON A POSSIBLE QUANTUM-LIKE BEHAVIOR FOR MENTAL STATES

Since the advent of quantum mechanics in 1927, there was a continuous debate on a possible connection between quantum and mental phenomena. It is really impossible to analyze in detail the various proposals that were prospected on this subject in the last seventy years and so we will briefly mention only a few of them.

Starting with 1929, A.N. Whitehead [22] made attempts to establish quantum and mental connections. A.N. Whitehead's formulation holds on an unusual statistical behavior of quantum systems and this model enabled this author to speculate that quantum systems represent cognitive systems in a generalized sense. Also A. Shimony [23] moved on a similar line of investigation.

According to the most accepted version of quantum mechanics, the so called Copenhagen interpretation, physical reality is created during the subjective act of its observation and measurement. In consequence of such principle in quantum theory, the problem of quantum-mental connection assumed basic relevance in the same problem of quantum measurement. Authors as E. Schrödinger [24], J. Von Neumann [25], W. Heisenberg [26], E.P Wigner [27], and, more recently, N.D. Mermin [28], A. Peres [29] and B. d'Espagnat [30] discussed the problem on this basis. The conclusion seemed to be that, according to the orthodox Copenhagen interpretation of quantum mechanics, the wave function provides a complete description of an individual quantum system. An act of observation or of measurement resolves itself in a collapse of wave function. Wave function collapse is the actual resolution of potential alternatives previously represented in the linear superposition of the wave function connected to the considered quantum system. According to various authors, e.g. [31], [32], the collapse of wave function connects directly an act of thinking. These authors suggested that the act of conscious thinking is itself the same of collapse resolving out potential alternatives, represented by the wave function, in final actualization. This idea of quantum physical reduction from potential alternatives to actualization as basic mechanism of cognitive processes became well accepted in several studies on application of quantum mechanics in mental dynamics. H. Stapp [33], in particular, considered a kind of reductionistic approach to the problem. He considered that brain processes involve in last analysis chemical processes that cannot escape to obey and to be treated quantum mechanically.

However, it seems to us that any reductionistic approach, finalized as general methodology to brain's study, cannot result so decisive in understanding cognitive processes. The sense of this position is cleared considering as example that new possibilities seemed to arise rather recently with the new potentialities to study neuron firings but it followed a strong disappointment in the conclusive possibility to approach some physical reduction of mental processes by this way. This is a reason to assume a strong critical position against the acceptance of a simple and reductionistic translation of quantum models in the sphere of cognitive dynamics. Another thing inducing non acceptance of automatic translation of quantum models in the sphere of cognition is that quantum micro-descriptions apply to a level of parameter description whose magnitude is far from magnitude of the corresponding brain parameters as temperature, time scale and so on.

The holistic quantum approach to cognitive dynamics was mainly based on Bohm-Hiley-Pylkkänen theory [34] of active information. These authors considered pilot waves as a kind of information field and they formulated interesting models of cognitive processes on this basis. Consciousness-information models were also developed by M. Lockwood and J.A. Barrett [35].
The actual nature of the problem results, however, more involved respect to a simplified modeling attitude often consisting in a quasi automatic translation of previously accepted physical foundations.

Starting with some years ago, one of us [20] conducted an articulated analysis of quantum formalism splitting the theory into two independent points, that one really quantum as the quanta, Planck constant and discreteness and the other independent part regarding the probabilistic formalism .It was evidenced that the probabilistic formalism, as it was considered by Born, Jordan, Hilbert, Dirac, von Neumann is a purely mathematical formalism that gives the possibility to characterize questions with regard to context dependent probabilities. In other terms, one has the possibility to consider not only purely and simple conditional probabilities but also probabilities depending on the complexes of the conditions,(contexts), respect to which events are actually realized. The relevant feature is that, by using straightforward frequency arguments, in [20] it was possible to classify transformations of probabilities which can be generated by transition from one preparation procedure to another, and this is to say from one context to another. Three classes of transformations were identified [20] and corresponding to statistical deviations of different magnitudes. We will be interested to trigonometric transformations corresponding to context-transitions that induce statistical deviations of relatively small magnitudes and that are negligible in the cases of phenomena supported from classical physics. The interesting case is that one of quantum-like violation of classical formula of total probability. It is well known that quantum violation of the classical formula of total probability, based on classical formula for conditional probabilities:

p(A= x) =p(B=+) p(A=x/B=+) + p(B=-) p(A=x/B=-),      (2.1)

where   x= +,-, with A and B dichotomous random variables assuming A=+,- and B = +, -, gives

p(A=x)=p(B=+)p(A=x/B=+)+p(B=--)p(A=x/B=-)
$$+ 2\sqrt{p(B=+)p(A=x/B=+)p(B=-)p(A=x/B=-)} \cos\vartheta(x)$$
(2.2)

where $\vartheta(+)$ and $\vartheta(-)$ are angles of phases, give measure of the happened trigonometric transformation. The obtaining of such transformation rule with the explanation of their direct connection with probabilities depending on the complexes of the contextual conditions, result to be of great importance. We say that systems following the (2.2) are context dependent and exhibit quantum-like behavior. We may apply such quantum-like formalism not only to the description of non deterministic micro phenomena but also to various phenomena outside the micro wold. One such possibility is to apply quantum-like formalism to describe statistical experiments with cognitive systems.

It must be clear that, following this line of investigation, we realize an approach that has no direct relation with reductionistic quantum models and we are not interested in statistical behavior of micro systems forming the macro system of the brain. We describe only probabilistic features of cognitive measurements. Our quantum-like approach describes statistics of measurements of cognitive systems with the aim to ascertain if cognitive systems behave as quantum-like systems where here quantum-like cognitive behavior means that cognitive systems result to be very sensitive to changes of the context with regard to the complex of the mental conditions.

## 3. DESCRIPTION OF THE MENTAL INTERFERENCE EXPERIMENT

The starting point of the experimentation is the thesis that quantum-like models may be useful to describe mental states. This is, in effect, the current hypothesis that we aim to ascertain. We have seen that the main distinguishing feature of quantum-like behaviors as compared to classical, deterministic one, is that in quantum like indeterministic pattern we are in presence of interference terms of probabilities that are due to the contextualized nature of the dynamics under consideration. In classical, statistical, deterministic physics the total probability of the event B= +, - is the sum of the single event probabilities. In quantum like non deterministic dynamics a strong dependence from the context act and consequently an interference term appears in the reckoning total probability. If such an interference term should be found in measurements of mental observables than such result should be interpreted as evidence in favor of the use of a quantum like approach in the description of mental measurements and in the same process of formation of mental functions.

Let A = +,- and B = +,- be two dichotomous mental observables. They are assumed to be two different kinds of cognitive tasks. We prepare an ensemble E of cognitive systems, that is to say, we select an ensemble of human beings having the same mental state. Then, on E we perform measurements of A =+,- and we obtain ensemble probabilities

p(A=+) and p(A=-) = 1 – p(A=+)

so that p(A=+) is the probability to get the result (+) under the measurement on cognitive system belonging to E, and similarly p(A=-) is the probability to get the result (-).

The same result may be obtained if we perform measurement by the dichotomous mental observable B= + , - obtaining P(B=+ ) and p(B = - ) = 1 – p( B=+) .

The next step is to prepare two ensembles $E_1$ and $E_2$ of cognitive systems having the states corresponding to B=+ and B = - . We may perform now the measurement of the A mental observable and thus obtaining

p(A=+/B=+) ; p(A=- / B= +) ; p(A=+/B=-) ; p(A=- /B = -).

P(A= i /B = j) represents the probability to obtain A = i (i = +,- ) having previously obtained B = j (j=+,-). The previously mentioned classical probability theory tell us that all these probabilities result to be connected by the well known formula of total probability (2.1) while instead, in the case of quantum like behavior for such explored mental states, we will have that the quantum-like formula of total probability, the (2.2) , must hold. In this second case, in opposition to the previous classical one, we will conclude that the cognitive process employed in the dynamics of formation of the mental states in the case of the ensemble E ,submitted to measurements of mental observables A=+, - and B =+, - is a context dependent (mental) dynamics that consequently accords to quantum like behavior.

Let us explain the experiment in detail. It is well known that, starting with 1912 [36], Gestalt moved a devastating attack against the structuralism formulations of perception in psychology. Classical structuralism theory of perception was based on a reductionistic and mechanicistic conception that was assumed to regulate the mechanism of perception. There exists for any perception a set of elementary defining features that are at the same time singly necessary and jointly sufficient in order to characterize perception also in cases of more complex conditions. Gestalt approach introduced instead an holistic new approach showing that the whole perception behavior of complex images never results to be reduced to the simple identification and sum of elementary defining features defined in the framework of our experience.

During 1920's and 1930's Gestalt psychology dominated the study of perception. Its aim was to identify the natural units of perception explaining it in a revised picture on the manner in which the nervous system works. Gestalt main contributions have provide to day some understanding elements of perception through the systematic investigation of some fascinating features as the causes of optical illusions, the manner in which the space around an object is involved in the perception of the object itself, and, finally, the manner in which ambiguity plays a role in the identification of the basic laws of the perception. In particular, Gestalt psychology also gave important contributions on the question to establish how it is that sometimes we see movements when the object we are looking at, is not really moving. As we know , when we look at something we never see just the thing we look at. We see it in relation to its surroundings. An object is seen against its background . In each case we distinguish between the figure , the object or the shape and the space surrounding it that we call background or ground. The psychologist E. Rubin [37] was the first to systematically investigate this phenomenon, and he found that it was possible to identify any well-marked area of the visual field as the figure and leaving the rest as the ground. However, there are cases in which the figure and the ground may fluctuate and one is faced to consider the dark part as the figure and the light part as the ground, and viceversa. Only a probabilistic answer may be given on a selected set of subjects that will tend to respond on the basis of subjective and context dependent factors. The importance of figure-ground relationship lies in the fact that this early work of Rubin represented the starting point from which the Gestalt psychologists began to explain what to day are known as the organizing principles of perception. A number of organizing or grouping principles emerged from such studies of ambiguous stimuli. Three identified principles may be expressed as similarity, closure and proximity. Gestalt psychologists attempted to extend their work also at a more physiological level postulating an existing strong connection between the sphere of the experience and the physiology of the system by admitting the well known principle of isomorphism. This principle establishes that the subjective experience of human being as well as the corresponding nervous event have substantially the same similar structure. Rather recently [38], A. Keil, M. Müller, W.J. Ray, T. Gruber, T. Elbert presented evidence of a neural correlation of the differential treatment of figure and ground by the brain. The effect of figure/ground assignment was observed even in the earliest portions of cell response, suggesting an intimate coupling between shape selectivity and figure/ground segregation. These new physiological findings resulted in a satisfactory accord with the perceptual effects that were described by E. Rubin.

In the present paper we examined subjects by tests A and B in order to test quantum like behavior as predicted by the (2.2).For test A and B we used the ambiguity figures of Fig. 1 as they were largely employed in Gestalt studies. The reasons to use such ambiguity tests here to analyze quantum like behavior in perception, may be summarized as it follows. First of all, the Gestalt approach was based on the fundamental acknowledgement of the importance of the context in the mechanism of perception. Quantum like behavior formulates the same basic importance and role of the context in the evolution of the considered mechanism. Finally, we have seen that in ambiguity tests, the figure and the ground may fluctuate during the mechanism of perception. Thus, consequently, a non deterministic (and this is to say …a quantum like) behavior should be involved.

## 4. MATERIALS AND METHODS

Ninety eight students in medicine with about equal distribution of females and males, aged between 19 and 22 years, from the University of Bari, Italy, provided informed consent to participate in the experiment.. In the first experiment a group of fifty three students was subjected in part to test A ( presentation of only test A) and in part to tests B and A (presentation of test B and soon after presentation of test A with prefixed time separation of about two seconds between the two tests). The same procedure was employed in the second and the third experiments for groups of twenty four and twenty one students, respectively. All the students of each group were submitted simultaneously to test A or to test B followed by test A. Ambiguity figures of tests A or B followed by A appeared on a large screen by a time of only three seconds and simultaneously it was asked to students to bar on a previously prepared personal schedule their decision to retain figures to be equal or not. Submission to students of test B soon after followed by test A had the finality to evaluate as the perception of the first image (test B) could alter perception of the subsequent image (test A). All the experiment was computer assisted and in each phase of the experimentation the following probabilities were calculated
$p(A=+)$, $p(A=-)$, $p(B=+)$, $p(B=-)$, $p(A=+/B=+)$, $p(A=-/B=+)$, $p(A=+/B=-)$, $p(A=-/B=-)$
and than a statistical analysis of the results was performed in order to ascertain if the (2.1) or instead the (2.2) occurred in the case of our measurements of mental observables as they were performed by us with Tests A, B, and A/B.

## 5. RESULTS AND CONCLUSIONS

The first experimentation gave the following results
Test A : p(A=+) = 0.6923 ; p(A=-) = 0.3077, (5.1)
Test B : p(B=+) =0.9259 ; p(B=-) = 0.0741 ;
Test A/B : p(A=+/B=+) = 0.68 ; p(A=-/B=+) =0.32 ,
p(A=+/B=-) =0.5; p(A=-/B=-) =0.5
The calculation of conditional probability gave the following result with regard to p(A=+)
p(B=+) p(A=+/B=+) + p(B=-) p(A=+/B=-) = 0.6666 (5.4)
The second experimentation gave the following results
Test A : p(A=+) = 0.5714 ; p(A=-) = 0.4286 , (5.5)
Test B: p(B=+) = 1.0000 ; p(B=-) =0.0000 ;
Test A/B : p(A=+/B=+) = 0.7000 ; p(A=-/B=+)0.3000 ;
p(A=+/B/-) = 0.000 ; p(A=-/B=-) = 0.0000 .
The calculation of the conditional probability gave the following result with regard to p(A=+)
p(B=+)p(A=+/B=+) + p(B=-) p(A=+/B=-) = 0.7 (5.8)
Finally, the third experimentation gave the following results
Test A : p(A=+) =0.4545 ; p(A=-) = 0.5455 , (5.9)
Test B: p(B=+) =0.7000 ; p(B=-) = 0.3000 ;
Test A/B : p(A=+/B=+) = 0.4286 ; p(A=-/B=+) = 0.5714; p(A=+/B=-) = 1 , p(A=-/B=-) = 0
The calculation of the conditional probability with regard to p(A=+) gave the following result
P(B=+) p(A=+/B=+) + p(B=-) p(A=+/B=-) = 0.6000 (5.12)
We underline that contextual scenario took place not in the physical space (Euclidean plane) where the geometric images were placed, but in the mental space. In the mental space the preceding appearance of the B-image disturbed the mental state S of the students and the successive appearance of the A-image was percept by students having a disturbed mental state.

The results are reported in Table 1 where statistical analysis is also included. It is seen that the mean value of p(A=+) resulted to be p(A=+) = 0.5727 ± 0.1189 with regard to the Test A and calculated by using the (5.1), the (5.5), and the (5.9) while instead a mean value of 0.6556 ± 0.0509 resulted for p(A=+) when calculated with regard to the Test A/B and thus using the (5.4), the (5.8), and the (5.12) . The two obtained mean values result to be different and thus give evidence for the presence of quantum like behavior in measurements of cognitive mental states as they were performed by testing mental observables by Tests A, B, and A/B .In order to confirm such difference, we used Student's t-Test.. As it is well known, if we have a collection of samples, it may arise the problem to estimate if the measured difference in their averages is large enough to reject the null hypothesis that in fact such differences are due to chance. Student's T-Test may be applied in these cases. It was performed on the set of data that we obtained in the course of the experimentation and it gave as result that we had no more than 0.30 probability that the obtained differences between the two estimated values of p(A=+) by Test A and by Test A/B (see the (2.1) and the (2.2)), were produced by chance.

It is evident that we need still more cases of experimentation in order to arrive to a more definitive quantification of the processes that we had under examination and to analyze definitively the thesis of quantum like behavior during cognitive measurements. However, such preliminar experimentation gave a first substantial evidence on the presence of quantum like behavior in cognitive dynamics of mental states. In accord with such result the conceptualizing function must be retained as strong dependent from the dynamics of the acting context.

As final step we may proceed now to calculate $\cos\vartheta(x)$ as given in the (2.2) and representing, as we know, a trigonometric measure of transformation of probabilities that are generated by transition from one context (the case of Test A) to another context (the case of Test A/B) . We call $\vartheta$ an angle of phase and its value more near to the basic values $(k+1) \pi/2$ $(k=0,1,2,..)$ will indicate a very modest influence of the context on the probabilities of transition from one context to another, while values of $\vartheta$ near to $2k\pi$ will indicate instead a strong role of the context in determining transition of probabilities. In the case of our experimentation we obtained $\cos\vartheta(+) = -0.2285$, $\vartheta(+) = 1.8013$ and $\cos\vartheta(-) = 0.0438$, $\vartheta(-) = 1.527$ that are quite satisfactory phase results in order to admit quantum like behavior for the investigated mental states.

As final step, we may now proceed by a detailed calculation of the quantum like model of the mental states as they were characterized during the experimentation

By using the obtained data, we can write a quantum-like wave function $\phi = \phi_S$ of the mental state S (of the group of students participated in the experiment). The general formula for representation by the complex amplitude was proposed in [20] :

$$\phi(x) = \sqrt{P(B=+) \, P(A = x / B = +)} + e^{i\theta(x)} \sqrt{P(B=-) \, P(A = x / B = -)}$$
(5.13)

The $\phi$ is a function from the range of values $\{+, -\}$ of the mental observable A to the field of complex numbers. Since the A may assume only two values, such function can be represented by two dimensional vectors with complex coordinates. We remark that we have Born's probability rule

$$|\phi(\pm)|^2 = P(A = \pm) \qquad (5.14)$$

Our experimental data give

$$\phi(+) = \sqrt{0.8753 \times 0.6029} + e^{i\theta(+)} \sqrt{0.1247 \times 0.5} \approx 0.7193 + i\, 0.2431$$
(5.15)

and

$$\phi(-) = \sqrt{0.8753 \times 0.3971} + e^{i\theta(-)} \sqrt{0.1247 \times 0.5} \approx 0.5999 + i\, 0.2494$$
(5.16)

By performing the same experiment for every group of people having some special mental state G we can calculate the wave function $\phi_G$ giving a quantum-like representation of the G. Thus there exists the map J mapping mental states into quantum-like wave functions. Such a mapping provides a mathematical representation of mental functions. A mental state is too complex object to provide its complete mathematical description, but we can formalize mathematically some features of a mental state by using the quantum-like representation. We remark that every quantum-like representation induce the huge reduction of information. In particular, the wave function $\phi_S$ calculated in our experiment gives a very rough representation of the mental state S. The $\phi_S$ contains just information on ability of students to percept contexts on the A and B-pictures.

On the space of complex functions we introduce the structure of a Hilbert space H with the aid of the scalar product

$$(\phi, \psi) = \phi(+) \overline{\psi}(+) + \phi(-) \overline{\psi}(-) \quad\quad\quad (5.17)$$

Thus J maps the set of mental states into the H. The mental observable A can be represented by the multiplication operator in H:

$$\hat{A} \phi(x) = x \, \phi(x) \; ; \; x = \pm \quad\quad\quad (5.18)$$

We see [20] that the mean value of the mental observable A in the mental state S can be calculated by using the Hilbert space representation

$$E_S A = (\hat{A} \phi, \phi) \quad\quad\quad (5.19)$$

In our concrete experiment by using experimental data, we have

$$E_S A = P(A = +) - P(A = -) = 0.1454 \quad\quad\quad (5.20)$$

The same result gives our quantum-like model

$$(\hat{A}\phi_S, \phi_S) = |\sqrt{0.8753 \times 0.6029} + e^{i\theta(+)}\sqrt{0.1247 \times 0.5}|^2 - |\sqrt{0.8753 \times 0.3971} + e^{i\theta(-)}\sqrt{0.1247 \times 0.5}|^2 = 0.1454 \quad (5.21)$$

We underline that the presented quantum-like representation of the mental state S is based on two concrete mental observables (reference observables) A and B; the mental wave function depends on such reference observables. In quantum physics the role of reference variables play the position and momentum observables [20]. However, in quantum-like approach we do not fix the reference observables for ever. Thus the same mental state can have different quantum-like representations based on different reference observables.

In conclusion, the experimental results presented in this paper support evidence of quantum-like statistical behavior of cognitive systems, at least human beings, with indication of mental phenomena not embedded in the traditional physical space. The consequent elaboration of the corresponding quantum like model has given the possibility to represent directly for the first time mental states on the basis of such quantum like model by Hilbert space vectors and complex amplitudes. Such a representation induces huge reduction of information about a full representation of a mental state.

**Table 1. Calculated probabilities following measurements of mental observables performed by Tests A, B, A/B**

| Experiments | I | II | III | IV | V | VI | VII |
|---|---|---|---|---|---|---|---|
| | p(A=+) | p(A=-) | p(B=+) | p(B=-) | p(A=+/B=+) | p(A=+/B=-) | p(B=+)p(A=+/B=+) + p(B=-) p(A= +/B =-) |
| 1 | 0,6923 | 0,3077 | 0,9259 | 0,0741 | 0,6800 | 0,5000 | 0,6667 |
| 2 | 0,5714 | 0,4286 | 1,0000 | 0,0000 | 0,7000 | 0,0000 | 0,7000 |
| 3 | 0,4545 | 0,5455 | 0,7000 | 0,3000 | 0,4286 | 1,0000 | 0,6000 |
| Mean Value | 0,5727 | 0,4273 | 0,8753 | 0,1247 | 0,6029 | 0,5000 | 0,6556 |
| Standard Deviation | 0,1189 | 0,1189 | 0,1563 | 0,1563 | 0,1513 | 0,5000 | 0,0509 |

**Student's t-Test: Results for groups I and VII.**

t = -1,1100 and st.dev.= 0,0915
The probability of this result, assuming the null hypothesis, is 0,3300

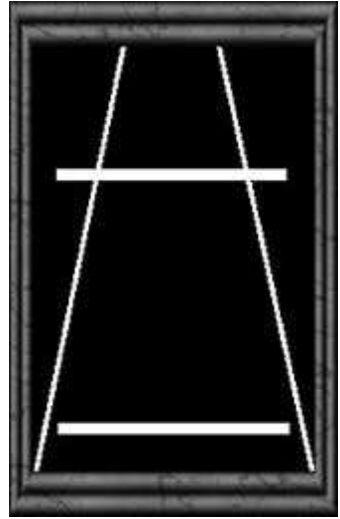 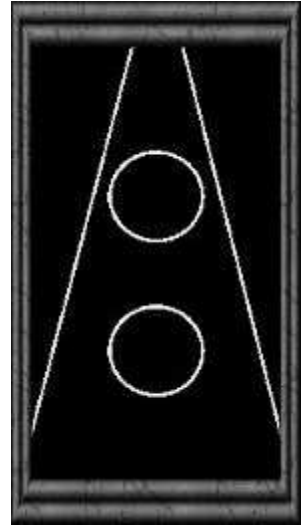

Fig.1          Test A                              Test B